\documentclass[page-classic]{epl2}%
\usepackage{graphics}
\usepackage{graphicx}
\usepackage{dcolumn}
\usepackage{amsmath}
\begin{document}
\title{A ``metaphorical'' nonlinear multimode fiber laser approach to weakly-dissipative Bose-Einstein condensates}
\shorttitle{A ``metaphorical'' nonlinear multimode fiber laser approach}
\author{V. L. Kalashnikov\inst{1,2} \and S. Wabnitz\inst{1,3}}

\shortauthor{V. L. Kalashnikov and S. Wabnitz}
%
%
\institute{ \inst{1} Dipartimento di Ingegneria dell'Informazione, Elettronica e Telecomunicazioni, Sapienza Universit\`a di Roma, via Eudossiana 18, 00184 Rome, Italy \\
 \inst{2} Institute of Photonics, Vienna University of
Technology, Gusshausstrasse 25-29, Vienna 1040, Austria \\
 \inst{3} Novosibirsk State University, Pirogova 1, Novosibirsk
630090, Russia}
\date{Received: date / Revised version: date}
%


\abstract{
We demonstrate the stabilization of two-dimensional nonlinear wave patterns by means of a dissipative confinement potential. Our analytical and numerical analysis, based on the generalized dissipative Gross-Pitaevskii equation, makes use of the close analogy between the dynamics of a Bose-Einstein condensate and that of mode-locked fiber laser, operating in the anomalous dispersion regime. In the last case, the formation of stable two-dimensional patterns corresponds to spatiotemporal mode-locking, using dissipation-enhanced mode cleaning. We analyze the main scenarios of pattern destabilization, varying from soliton dissolution to its splitting and spatiotemporal turbulence, and their dependence on graded dissipation.} 
%
\maketitle
\section{Introduction}
\label{intro}

Emerging phenomena comprising turbulence, light, matter and quasi-particle Bose-Einstein condensation, including the transition between coherent and non-coherent collective states mediated by spatio-temporal turbulence \cite{RefBEC,RMP,RefCondensation,light,polar,RefTurbulence,k1,zakharov}, bridge macro- and quantum physics, and open the way to a ``mesoscopic'' quantum world \cite{M}. An impressive example of coherent quantum mesoscopic state is provided by Bose-Einstein condensation (BEC). BEC can be treated as a matter-wave soliton, similar to the soliton solution of the nonlinear Schr\"{o}dinger equation (NSE) in optics \cite{BECS,solitons}. Such analogy is based on the equivalence of the underlying one-dimensional (1D) models, where time coordinate, kinetic energy, and attractive interaction of bosons for a BEC, correspond to propagation distance, anomalous group-dispersion (or diffraction for spatial solitons \cite{spatial}), and self-phase modulation (self-focusing) in optics, respectively.

On the one hand, it is well-known that 2D solitons undergo a catastrophic collapse \cite{spatial,kelley}. On the other hand, it is also known that BEC solitons can be stabilized by means of introducing a trapping potential \cite{bredley,malomed1,malomed2}, or by tuning the effective inter-atom interaction potential from repulsive to attractive through a Feshbach resonance \cite{khaykovich}. In particular, this can be achieved with a  periodic temporal modulation of nonlinearity \cite{abdullaev,saito,montesinosa,adhikari,malomed3}, or by means of a modulation of the trapping potential \cite{malomed3}. A whole zoo of different well-localized spatial coherent and semi-coherent states with non-zero topological charges was theoretically demonstrated (see \cite{malomed1} for an overview).

A similar stabilization methodology for ultra-short optical pulse propagation is essential in optical telecommunications, where the confinement potential can be introduced by means of a graded refractive index (GRIN) in optical fibers \cite{agrawal1}. Moreover, the possibility of a periodical modulation of the group-velocity dispersion (GVD) in optical fiber systems, and the use of different nonlinearity mechanisms, may provide additional tools for spatiotemporal soliton stabilization \cite{malomed3,ss,malomed4}. The close analogy between nonlinear phenomena in atomic and photonic systems makes the latter an ideal test-bed for exploring fundamental physics, ranging from plasma to BEC \cite{metap}.

An additional advantage of the analogy between BEC and nonlinear optical systems concerns the scaling of the boson amount, which is the BEC mass or the optical soliton energy, respectively. This issue has two principal aspects. The first one is the contribution of dissipative effects to the properties of BEC \cite{d1,d2,d3} and optical solitons \cite{ds}. Such a contribution can be crucial in the process of coherent condensate (or soliton) self-emergence, and its stability \cite{malomed4}. As it was demonstrated, pulse manipulation with the help of a graded dissipation in a fiber laser could provide the means to achieve the self-emergence of stable dissipative solitons, by the so-called distributed Kerr-lens mode-locking technique \cite{klm}. Our first proposal is to expand this method to BEC.

The second aspect concerns the scaling of the condensate mass/energy, which can be provided by varying the condensate size. In terms of fiber optics, this means using GRIN multimode fibers (MMFs), where spatial instabilities could destabilize pulse dynamics. It was found that the effect of nonlinear spatial mode cleaning in MMFs can suppress such instabilities, and provide a way to achieve spatiotemporal mode-locking, i.e., the formation of localized stable coherent spatial and spatiotemporal patterns \cite{wise1,wabnitz1}.

In this Letter, we consider a generalized dissipative model based on the 2D Gross-Pitaevskii equation (GPE), taking into account the presence of graded dissipation, and mass/energy exchange with a non-coherent environment. The analytical soliton solution corresponding to the ground soliton state (i.e., fundamental mode soliton in photonics) is demonstrated. Furthermore, the stabilization of 2D coherent structures by graded dissipation (dissipative mode cleaning) is numerically analyzed.

\section{2D dissipative soliton}
\label{sec:1}
Let us consider a particular case of BEC, formed by an axisymmetric harmonic potential of a cigar type with confinement along $r-$axis, and unconfined along the $z-$axis. In photonics, this is a model for ultrashort pulse dynamics in a GRIN fiber laser (the $z-$coordinate corresponds to a local or ``retarded'' time coordinate in that case). The last statement is akin to a specific space-time duality in optical signal processing \cite{kolner,gaeta}. Thus, anomalous GVD plays the role of the $z-$component of the boson kinetic energy. The corresponding dimensionless master equation is the generalized 2D GPE:

\begin{equation} \label{GPE}
  i\frac{{\partial \psi}}{{\partial T}} = \frac{1}{2}\left[ {\frac{1}{r}\frac{\partial }{{\partial r}}r\frac{\partial }{{\partial r}} + \left( {1 + i\tau } \right)\frac{{{\partial ^2}}}{{\partial {z^2}}}} \right]\psi - \frac{1}{2}\left( {1 + 2i\kappa } \right){r^2}\psi + {\left| \psi \right|^2}\psi - i(\Lambda  - \sigma \int\limits_{ - \infty }^\infty  {{{\left| \psi \right|}^2}dz} )\psi.
\end{equation}

Here, axial symmetry and zero vorticity are assumed for the wave-function $\psi(T, r,z)$ (which is a local-time dependent field amplitude in photonics).  The first term defines the kinetic energy (diffraction/dispersion in photonics) with a ``kinetic cooling'' along the unconfined $z-$axis. The last results from the growth of escaping rate with kinetic energy (that is, spectral dissipation in photonics) and it is defined by the $\tau-$parameter. The second term describes the complex parabolic confining potential, whose imaginary part $\kappa$ contributes for relatively large electric fields \cite{d1,d3,d4}. In photonics, this term characterizes a GRIN fiber profile, with a graded dissipation tracing the parabolic refractive index profile. The third nonlinear term is related to the two-body-scattering length. We consider an attractive interaction in a condensate, which defines the sign before this term: this corresponds to self-phase modulation (self-focusing) in photonics. The last term describes nonlinear loss caused by a weak dissipative condensate-basin interaction. Here, $\Lambda<0$ corresponds to a net linear gain (a ``gain'' means an inflow from non-coherent ``basin'' to the condensate) \cite{d3}. We impose a saturation of such gain with the condensate mass growth, as defined by the $\sigma-$parameter. This effect is an analog of the gain saturation in a laser, and it causes the feedback reaction of an effective potential on the condensate transformation, which can be characterized by the effective potential width (effective ``aperture size'') parameter $\chi  = \sqrt {{{\left| {\Lambda  - \sigma \int {{{\left| \psi  \right|}^2}dz} } \right|} \mathord{\left/
 {\vphantom {{\left| {\Lambda  - \sigma \int {{{\left| \psi  \right|}^2}dz} } \right|} \kappa }} \right.
 \kern-\nulldelimiterspace} \kappa }} $.

The approximate solution of Eq. (\ref{GPE}) corresponding to the soliton-like ansatz

\begin{equation} \label{ansatz}
    \psi \left( {T,r,z} \right) = {\psi _0}(T)~{\mathop{\rm sech}\nolimits} \left( {\frac{z}{{{\rm Z}\left( T \right)}}} \right)\exp \left[ {i\left( {\phi \left( T \right) + \zeta \left( T \right){z^2} + \theta \left( z \right){r^2}} \right) - \frac{{{r^2}}}{{2{\rm P}{{\left( T \right)}^2}}}} \right],
\end{equation}
($\psi_0$, \rm P, \rm Z, $\zeta$, $\theta$, and $\phi$ are the time-dependent amplitude, spatial widths, chirp, and phase, respectively) can be obtained by using the variational approximation and the Kantorowitch's method \cite{VA,dklm}.

The $T-$independent solutions for the soliton parameters are:

\begin{eqnarray}
    \zeta = \frac{8 \sqrt{15} \tau }{\pi ^2 {\rm Z}^2 \left(\sqrt{15 +128 \tau ^2}+\sqrt{15} \right)},~ \theta  =  - \frac{1}{4}\left( {2\kappa \,{{\rm P}^2} + \sigma \psi _0^2{\rm Z}} \right),\nonumber \\
\psi_0^2 = \frac{6 \left(1-\rm P ^4-\kappa ^2 \rm P ^8\right)}{\rm P ^2 \left(\sqrt{9 \sigma ^2\left(1-\rm P ^4\right)  \rm Z^2+6 \kappa \sigma \rm Z \rm P ^4+1}+3 \kappa \sigma \rm Z \rm P ^4+1\right)}.
\end{eqnarray}

\noindent The solution for $\rm Z$ and the final equation for $\rm P^2$ are too cumbersome, and we do not explicitly show them here \cite{dklm}. The equation for $\rm P^2$ contains radicals and has to be solved numerically.

\begin{figure*}[htpb]
\begin{center}
\includegraphics[width=14cm]{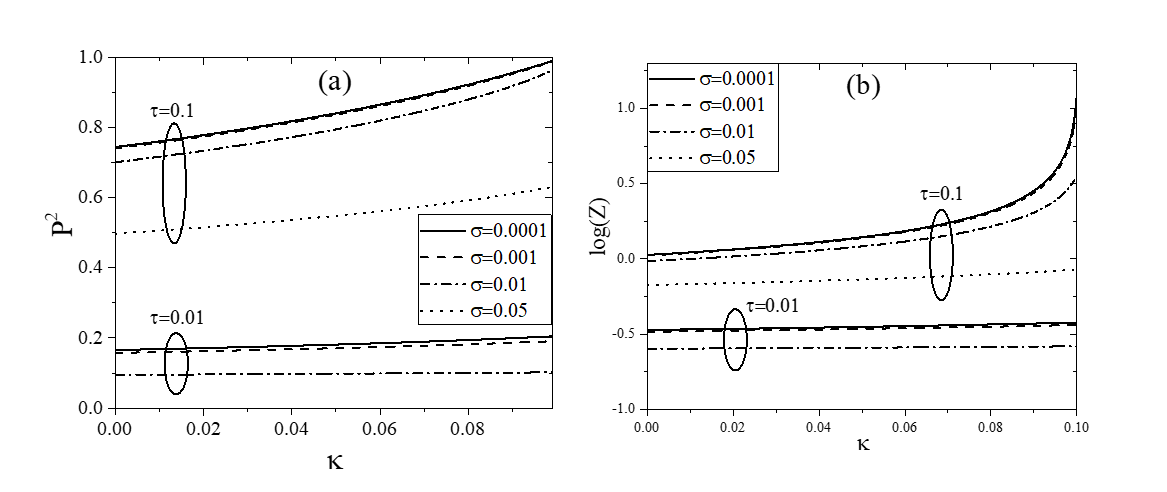}
\caption{(a): Transverse area of the soliton ($\rm P ^2$), and $\rm Z$ (b) as a function of the gain saturation parameter $\sigma$, the graded dissipation parameter $\kappa$ and the kinetic cooling parameter (spectral dissipation in photonics) $\tau$, $\Lambda=-0.1$.}
\label{fig:1}
\end{center}
\end{figure*}

\begin{figure*}[htpb]
\begin{center}
  \begin{minipage}{0.45\textwidth}
\includegraphics[width=6.2cm]{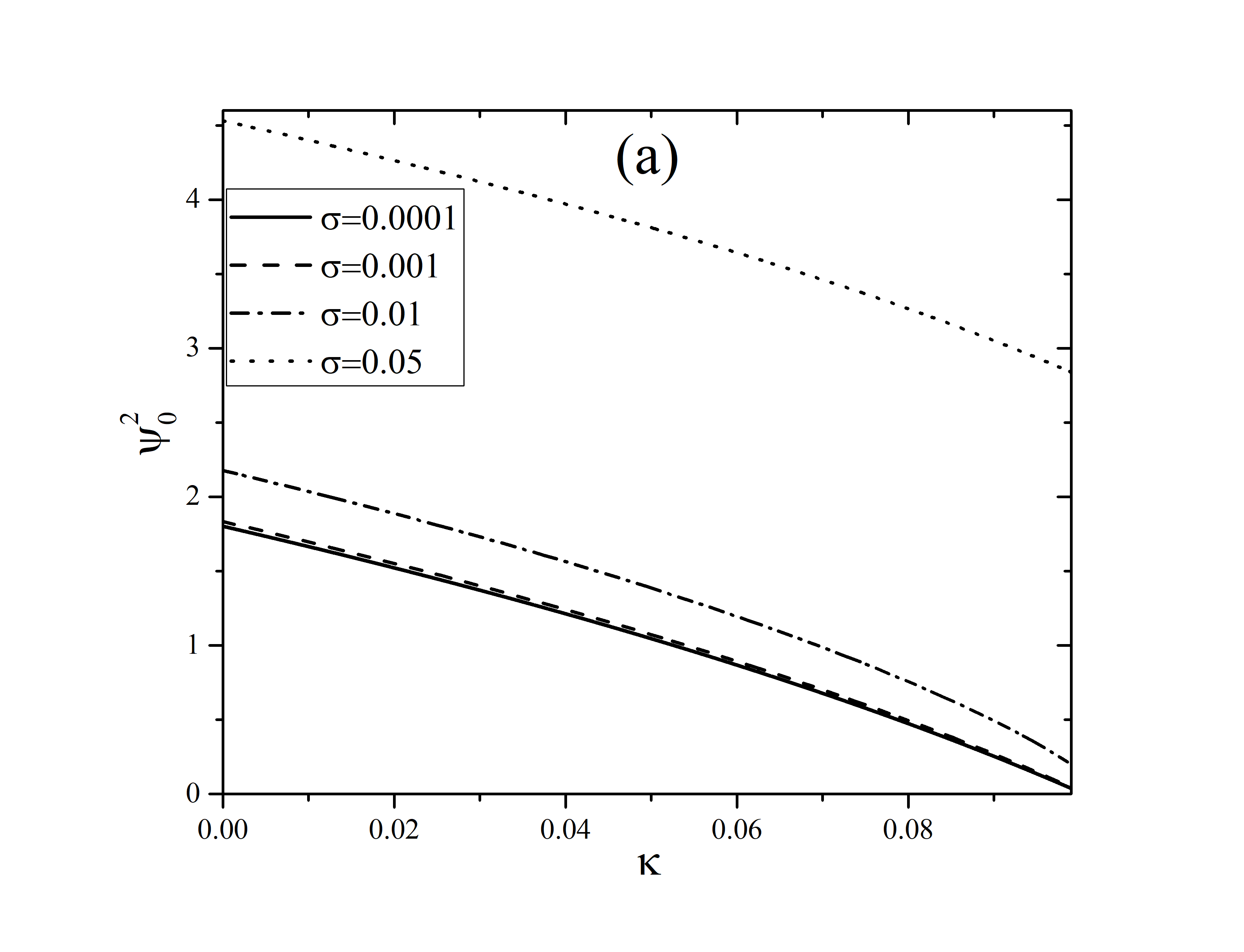}
  \end{minipage} \quad
  \begin{minipage}{0.45\textwidth}
  \includegraphics[width=6.2cm]{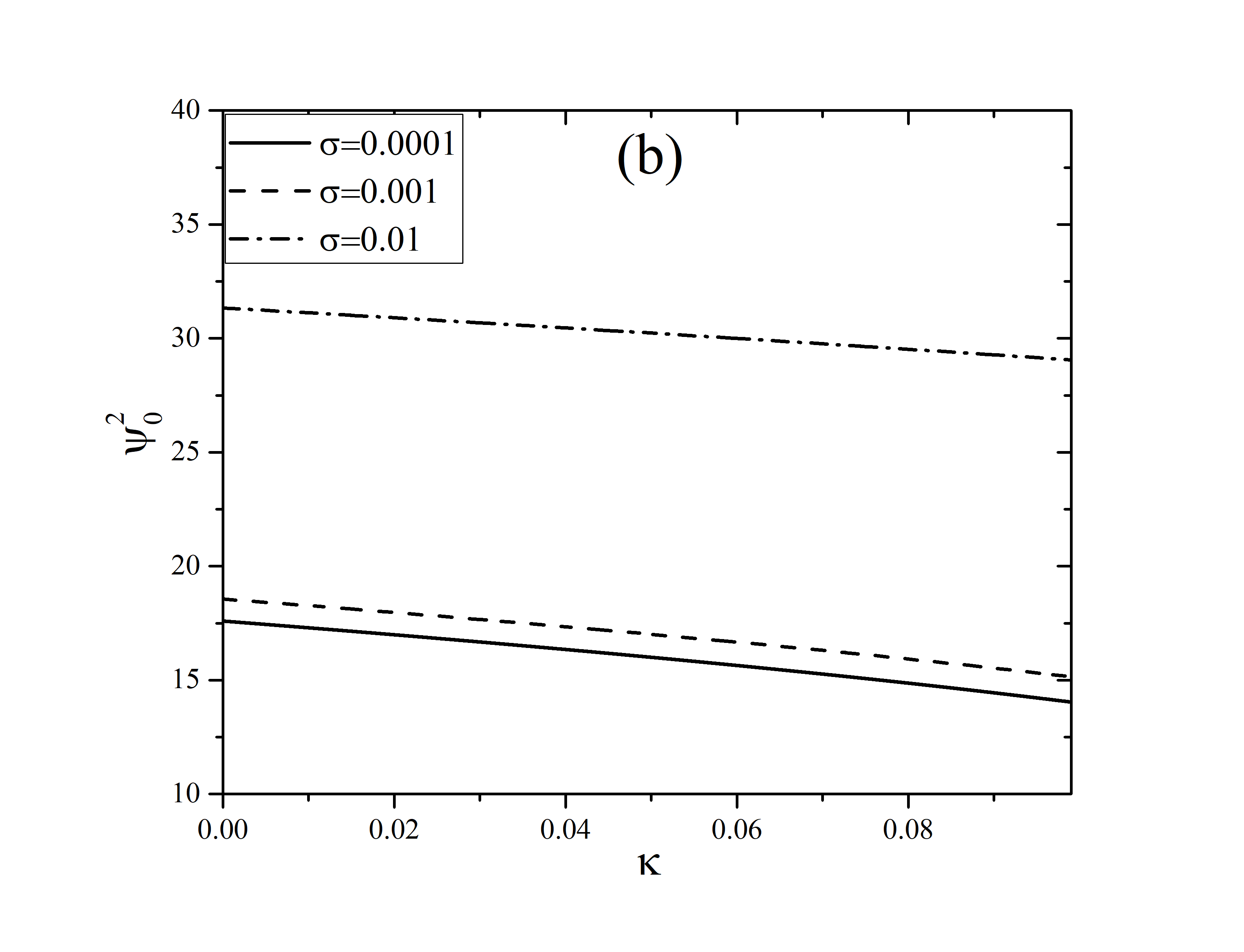}
\end{minipage} \quad
\caption{Squared soliton amplitude for $\tau=0.1$ (a), and $\tau=0.01$ (b) as a function of the graded dissipation parameter $\kappa$, and the gain saturation parameter $\sigma$. $\Lambda=-0.1$.}
\label{fig:3}
\end{center}
\end{figure*}

The dependence of soliton parameters on the various terms in Eq. (\ref{GPE}) is shown in Figures \ref{fig:1}-\ref{fig:3}. The specific values of our dimensionless parameters are close to those considered in \cite{d3}. Figs. \ref{fig:1} illustrate the dependence of the soliton spatial size on graded dissipation ($\kappa$), kinetic cooling ($\tau$), and gain saturation ($\sigma$), respectively. Gain saturation $\sigma$ leads to a decrease of the effective potential width $\chi$, which results in a spatial compression (or squeezing) of the soliton. This spatial squeezing is reduced as the graded potential $\kappa$ grows larger (see Figs. \ref{fig:1}). Correspondingly, the soliton amplitude $\psi_0^2$ decreases (see Fig. \ref{fig:3}). One can see from these Figures that a tenfold decrease of the kinetic cooling parameter $\tau$ leads to a significant increase of the soliton spatial squeezing, accompanied by a substantial growth of its amplitude.

However, we would like to point out that the results based on the ansatz of Eq. (\ref{ansatz}), i.e., the ground-state or fundamental mode assumption, are not conclusive, because they ignore the contribution of higher-order states (modes). Such a contribution can be grasped by numerical simulations of the dissipative 2D GPE.

\section{Dissipative mode-cleaning}
\label{sec:2}

For a specific model in optics that is analogous to BEC, we considered the master equation associated with spatiotemporal pulse propagation in a multimode fiber laser, based on active and nonlinear GRIN silica fibers. The graded dissipation in the fiber laser was characterized in terms of the previously discussed effective aperture size $\chi$. The presence of anomalous GVD, corresponding to a $z-$kinetic component, can be achieved by means of an appropriate waveguide compensation of the normal dispersion of silica when considering an active fiber dopant with gain centered around the carrier wavelength of 1.03 $\mu$m (e.g., ytterbium). For example, this can be obtained by the microstructuring of a fiber \cite{pcf}. Alternatively, the carrier wavelength can be shifted deeper into the infrared region, e.g., using an erbium-doped fiber at 1550 nm, in order to obtain anomalous material GVD. Following the analogy between BEC and photonics models, the longitudinal coordinate of the former corresponds to the local time coordinate $t$ of the latter. Whereas the GVD value defines the normalization of the $z-$coordinate for BEC. Let us recall that the time coordinate $T$ for BEC corresponds to the propagation distance along the optical fiber. We used the finite-element methods for the numerical simulations of Eq. (\ref{GPE}), under the condition of saturated gain, i.e., with $\Lambda \ll 1$.

In the absence of dissipation, a beam which is confined by an external potential (see Eq. (\ref{GPE})) exhibits a decay into a multitude of spatial patterns with complex dynamics (three spot-shoots are demonstrated in Fig. \ref{fig:4}). In a GRIN fiber, that corresponds to the presence of multimode beating or self-imaging \cite{wabnitz1}.

\begin{figure}[htpb] \label{fig:4}
\begin{center}
\includegraphics[width=12 cm]{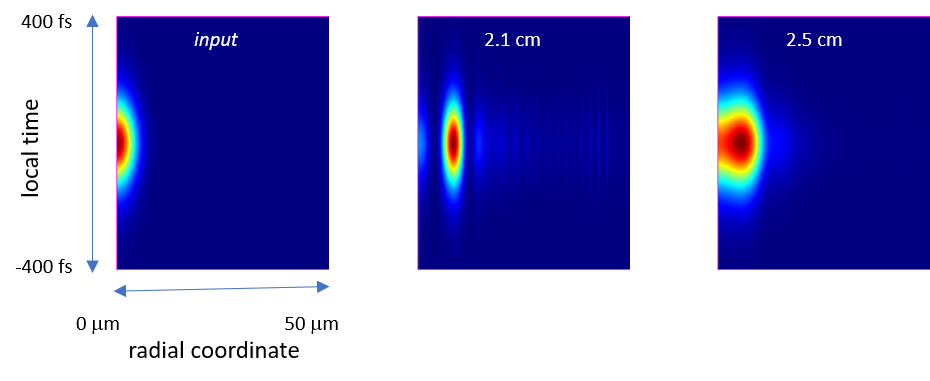}
\caption{\label{fig:4} Contour-plots of $|\psi|^2$ at three propagation distances in the absence of dissipation ($\Lambda=0$, $\kappa=0$). The following parameters characterize an MMF: the central wavelength is of 1.03 $\mu$m, the group-delay dispersion coefficient is of -0.022 fs$^2$/cm, the fiber core diameter is of 50 $\mu$m, and the confinement potential defined by the core/cladding refractive index difference is of 0.0103 (the cladding refractive index is of 1.457). The initial pulse with $\rm P=$7 $\mu$m, corresponding to 10 $\mu$m beam radius at $1/e-$level has a temporal width $\rm Z=$150 fs. The input peak power is of 100 kW, and the nonlinear refractive index coefficient, defining the power normalization, is of $2.7*10^{-8}$ $\mu$m$^2/$W.}
\end{center}
\end{figure}

The adjustment of graded dissipation provides mode-cleaning \cite{wabnitz1}: as shown by Fig. \ref{fig:5}, this is already effective for relatively short propagation distances, and moderate levels of $|\psi|^2$. This corresponds to the formation of a spatiotemporal soliton in the GRIN fiber \cite{wise2}. But we have to note that mode-cleaning and stable spatiotemporal soliton formation are only possible because of graded dissipation, including spectral filtering (i.e., kinetic cooling) \cite{wise1}. That directs to the mechanism of stable 2D-BEC formation by means of a manageable weak-dissipation.

\begin{figure}[htpb] \label{fig:5}
\begin{center}
\includegraphics[width=11 cm]{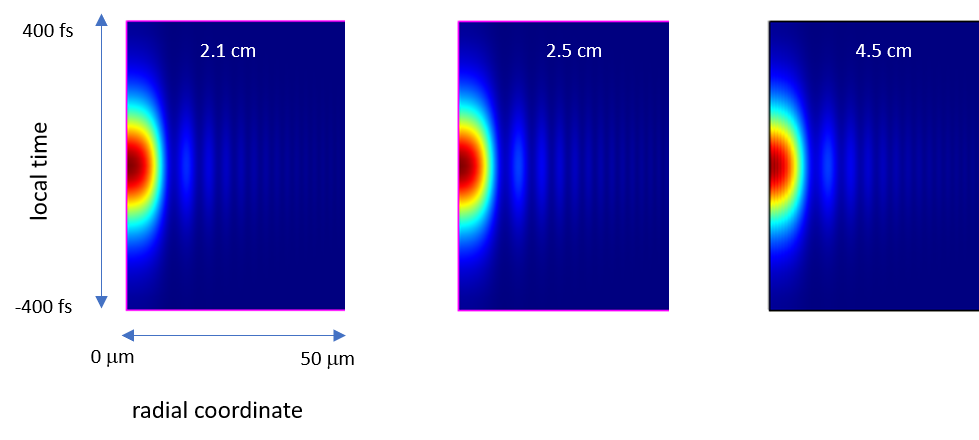}
\caption{\label{fig:5} Contour-plots of $|\psi|^2$ for $\Lambda=-0.001$, $\chi=37 \mu$m. $\tau$ is 1/100th of the GVD absolute value. Other parameters correspond to Fig. \ref{fig:4}.}
\end{center}
\end{figure}

Nevertheless, dissipation provokes additional destabilizing effects. The first one is rather trivial, and it involves soliton decay associated with a $\chi-$decrease. A more interesting effect is the observation of stable spatial-temporal fragmentation, corresponding to multiple pulse train generation in a fiber laser (Fig. \ref{fig:6}, (a)). This effect results from the decrease of graded dissipation (i.e., the $\chi-$parameter grows), so that the onset of dissipative mode-cleaning is hampered.

The analysis of the underlying dynamics demonstrates that the excitation of higher-order spatial modes leads to energy leaking into them, so that single-pulse generation becomes energetically unfavorable. As a result, the beam tends to relax to a state which is characterized by multi-pulse generation \ref{fig:6}, (a). This is akin to dissipative soliton multipulsing, which occurs as a result of the decrease of self-amplitude modulation (i.e., $\chi-$increase in our case) \cite{k1}. Such a transition to multipulsing can be interpreted as an excitation of the dissipative solitons' ``internal modes'' \cite{k2}. A further weakening of graded dissipation results in spatiotemporal turbulence, which occurs whenever the coherence between the mode patterns is fully broken (Fig. \ref{fig:6}, (b)).

\begin{figure}[htpb] \label{fig:6}
\begin{center}
\includegraphics[width=9 cm]{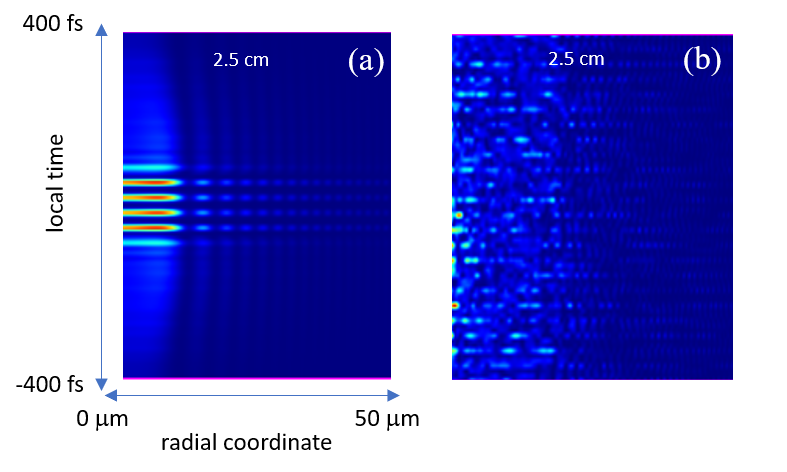}
\caption{\label{fig:6} Contour-plots of $|\psi|^2$ for: (a) $\chi=40 \mu$m, and (b) $\chi=50 \mu$m, $\tau$=1/50th of the GVD absolute value; $\Lambda=-0.001$. Other parameters correspond to Fig. \ref{fig:4}.}
\end{center}
\end{figure}

\section{Conclusion}
\label{sec:3}
In this work, we analyzed the stabilization of 2D patterns (solitons) in a weakly-dissipative BEC with a cigar-shaped confinement potential. The metaphorical modeling was based on the close analogy between nonlinear processes in 2D BEC and spatiotemporal soliton dynamics in a GRIN MMF. The underlying mathematical master equation is the generalized dissipative 2D+1 GPE, which was solved by the analytical but approximate variational approach, and by numerical methods. It was found that dissipative factors, such as 2D-graded dissipation, saturable gain, and kinetic cooling (spectral dissipation), play a crucial role in stabilizing spatiotemporal solitons. In photonics, one can treat such a stabilization as a manifestation of mode-cleaning in an MMF, which is enhanced by dissipation and can be realized on extremely short propagation distances. That could provide a spatiotemporal mode-locking mechanism in a fiber laser, akin to the regime of distributed Kerr-lens mode-locking in solid-state lasers.

We identified the main scenarios leading to the destabilization of coherent spatiotemporal nonlinear wave structures. The prevailing of graded dissipation, i.e., a too narrow effective aperture, causes 2D soliton dissolution. The growth of the aperture size results in the formation of stable 2D soliton. A further aperture widening causes a soliton splitting initially, i.e., the formation of 2D patterns, or spatiotemporal multipulse generation in photonics. When the contribution of graded dissipation becomes too weak, the condensate loses its coherence, and spatiotemporal turbulence develops. The main practical significance of the obtained results lies in the possibility of mass/energy scaling of coherent 2D matter wave structures, and the demonstration of stable spatiotemporal mode-locking in MMF lasers.

\acknowledgments
This work has received funding from the European Union Horizon 2020 research and innovation program under the European Research Council Advanced Grant No. 740355 (STEMS), the Marie Skłodowska-Curie Grant No. 713694 (MULTIPLY), and the Russian Ministry of Science and Education Grant No. 14.Y26.31.0017.

\end{document}